\documentclass[intlimits,twoside,a4paper]{article}
\usepackage[cp1251]{inputenc}

\usepackage[eqsecnum]{cmpj3}

\usepackage{bm}

\newcommand\tpp{t^{++}}
\newcommand\tpm{t^{+-}}
\newcommand\tmm{t^{--}}
\DeclareMathOperator{\Tr}{Tr}
\DeclareMathOperator{\adj}{adj}

\allowdisplaybreaks

\issue{2018}{21}{2}{23702}
\doinumber{10.5488/CMP.21.23702}

\title{Nonlocal correlations in the optical conductivity spectra}
\author{D.A. Dobushovskyi,
        A.M. Shvaika}
\address{Institute for Condensed Matter Physics of the National Academy of Sciences of Ukraine, \\ 1 Svientsitskii St., 79011 Lviv, Ukraine}

\date{Received May 2, 2018, in final form June 1, 2018}

\begin{document}

\maketitle

\begin{abstract} 
Optical conductivity spectra are studied for the Falicov-Kimball model with correlated hopping on the Bethe lattice. An expression for the current-current correlation function is derived using dynamical mean field theory. In the metallic phase with small correlated hopping values, the shape of Drude peak deviates from the Debye relaxation peak, and an additional wide peak is observed on the optical conductivity spectra and on Nyquist plot when Fermi level is in the vicinity of the two particle resonance. At  larger values of the correlated hopping parameter, the density of states contains three bands and the corresponding optical spectra and Nyquist plots display a more complicated shape with additional peaks. For strong local correlations, the correlated hopping reduces the width of the upper Hubbard band resulting in a decrease of the Drude peak spectral weight for the doped Mott insulator.
\pacs 78.20.Bh, 71.10.Fd, 71.27.+a
\end{abstract}

\section{Introduction}

Electron correlations attract a great interest in connection with various phenomena in different materials, from the one- and two-dimensional organic conductors, through three-dimensional solids, up to the optical lattices.  
As it was noticed by Hubbard in his seminal article~\cite{hubbard:238}, the second quantized representation of the inter-electron Coulomb interaction should also take into account, besides the local term $U\sum_i n_{i\uparrow}n_{i\downarrow}$, the nonlocal contributions including the inter-site Coulomb interaction $\sum_{ij} V_{ij} \hat{n}_{i}\hat{n}_{j}$ and the so-called correlated hopping
\begin{equation}
\sum_{ij\sigma} t_{ij}^{(2)} (\hat{n}_{i\bar{\sigma}}+\hat{n}_{j\bar{\sigma}})c_{i\sigma}^{\dagger}c_{j\sigma},
\end{equation}
which reflects the fact that the value of inter-site hopping depends on the occupation of these states. 
Correlated hopping can arise either due to a direct inter-site interaction or due to an indirect effective interaction~\cite{foglio:4554,simon:7471}.

Local Coulomb interaction has been widely investigated in the theory of strongly correlated electron systems, whereas the correlated hopping attracted much less attention.  It was mainly considered while elaborating new mechanisms for high temperature superconductivity~\cite{hirsch:326,bulka:10303}, and while describing organic compounds~\cite{arrachea:1173} and molecular crystals~\cite{tsuchiizu:044519}, electron-hole asymmetry~\cite{didukh:7893} and enhancement of magnetic properties~\cite{kollar:045107}.
Quite recently, the correlated hopping turned out to be an important puzzle in understanding the  behaviour of quantum 
dots~\cite{meir:196802,hubsch:196401,tooski:055001} and fermionic~\cite{jurgensen:043623,liberto:013624} and 
bosonic~\cite{eckholt:093028,luhmann:033021,jurgensen:193003} atoms on optical lattices.

Here, we study the effects of correlated hopping using the Falicov-Kimball model~\cite{falicov:997}, the simplest model  of strongly correlated electrons, which considers the local interaction between the itinerant $d$ electrons and localized $f$ electrons.
It is a binary alloy type model and its ground state phase diagram for the one-dimensional $(D=1)$ 
and two-dimensional $(D=2)$ cases displays a variety of modulated 
phases~\cite{gajek:3473,wojtkiewicz:233103,wojtkiewicz:3467,cencarikova:42701}.
It possesses an exact solution in infinite dimensions~\cite{brandt:365,freericks:1333} within the dynamical mean field theory (DMFT)~\cite{metzner:324,georges:13}.
An extension of the model by inclusion of correlated hopping was also considered, 
and the DMFT solution with a nonlocal self-energy was obtained~\cite{schiller:15660,shvaika:075101}.

In our previous article~\cite{dobushovskyi:125133}, we  considered a charge and heat transport in the Falicov-Kimball model with correlated hopping on Bethe lattice. Exact solutions for the one-particle density of states (DOS) and two-particle transport function (the ``quasiparticle'' scattering time) were obtained, and for a wide range of the correlated hopping, the singularities due to the resonant two-particle contributions were observed on a transport function, whereas the one particle DOS does not show any anomalous features. 
By tuning the doping of itinerant electrons, in order to bring the chemical potential close to the resonant frequency, a large increase of the electrical and thermal conductivities and of the thermoelectric power can be achieved. 
On the other hand, for some values of correlated hopping, when the hopping amplitude between the occupied sites is sufficiently reduced, itinerant electrons localize in the clusters of sites occupied by $f$-electrons giving rise to an additional narrow band in the DOS between the lower and upper Hubbard bands. 

The main subject of this article is to study to what extent the anomalous features observed in the DOS and in the transport properties of the Falicov-Kimball model with correlated hopping can be manifested in the dynamical response, such as optical conductivity.

The paper is organized as follows. 
In section~\ref{sec:formalism}, we present the DMFT solution to the Falicov-Kimball model with correlated hopping on a Bethe lattice. 
Section~\ref{sec:optcond} provides the derivation of the current-current correlation function and optical conductivity in a homogeneous phase. 
In section~\ref{sec:results}, we consider the peculiarities of the current-current correlation function and its Nyquist plots and of the optical conductivity for different values of the correlated hopping and doping. The results are summarized in section~\ref{sec:conclusions}.

\section{The model Hamiltonian}\label{sec:formalism}

The Falicov-Kimball model~\cite{falicov:997} with correlated hopping is described by the Hamiltonian
\begin{eqnarray}
&&H= H_{\textrm{loc}} + H_t\,,
\nonumber\\
&&H_{\textrm{loc}} = \sum_i \left[Un_{id}n_{if} - \mu_f n_{if} - \mu_d n_{id}\right],
\nonumber\\
&&H_t=-\sum_{\langle ij\rangle} \frac{t_{ij}^{*}}{\sqrt{Z}} \Bigl[t_1d_i^{\dag}d_j +
t_2d_i^{\dag}d_j\left(n_{if}+n_{jf}\right) 
+ t_3d_i^{\dag}d_jn_{if}n_{jf}\Bigr],
\end{eqnarray}
where $H_{\textrm{loc}}$ describes local interaction between the itinerant $d$-electrons and localized $f$-electrons, 
and $H_{t}$ represents the nonlocal hopping terms on the Bethe lattice with infinite coordination number, $Z\rightarrow\infty$, 
including the nearest-neighbor inter-site hopping $t_1$ and correlated hopping terms $t_2$ and $t_3$.
It is convenient to introduce the projection operators $P_i^+=n_{if}$ and $P_i^-=1-n_{if}$ and  rewrite
the nonlocal term in a matrix form~\cite{shvaika:075101},
\begin{align}
H_t&=-\sum_{\langle ij\rangle} \frac{t_{ij}^{*}}{\sqrt{Z}} \Bigl[
t^{++}P_i^+d_i^{\dag}d_jP_j^+ + t^{--}P_i^-d_i^{\dag}d_jP_j^-
+t^{+-}P_i^+d_i^{\dag}d_jP_j^- +
t^{-+}P_i^-d_i^{\dag}d_jP_j^+\Bigr]
\nonumber\\
&= -\sum_{\langle ij\rangle} \frac{t_{ij}^{*}}{\sqrt{Z}} \bm{d}_i^{\dag}\mathbf{t}\bm{d}_j\,,
\label{eq:Ht_mtrx}
\end{align}
where we have introduced a vector of the projected $d$-electron operators and a hopping matrix
\begin{equation}
\bm{d}_i=\begin{pmatrix} d_i P_i^+ \\ d_i P_i^- \end{pmatrix}, \qquad
\mathbf{t}=\begin{bmatrix}
t^{++} & t^{+-} \\
t^{-+} & t^{--}
\end{bmatrix} 
\label{eq:t_mtrx}
\end{equation}
and the hopping matrix elements read
\begin{equation}
t^{--}=t_1\,,  \qquad
t^{+-}=t^{-+}=t_1+t_2\,,  \qquad
t^{++}=t_1+2t_2+t_3\,.   
\end{equation}

\subsection{The single particle Green's function}

The Green's function for the projected $d$-electrons is defined by the matrix 
$\mathbf{G}_{ij}=[ {G}_{ij}^{\alpha\beta}]$, where ${\alpha,\beta=\pm} $. 
On the imaginary time-axis, we have 
\begin{equation}
\mathbf{G}_{ij}(\tau-\tau') = -\left\langle \mathcal{T} \bm{d}_i(\tau) \otimes \bm{d}_j^{\dagger}(\tau')\right\rangle, 
\end{equation}
where $\mathcal{T} $ is the imaginary-time ordering operator, $ \otimes $ denotes the direct (Cartesian) product of two vectors,
and the angular bracket denotes the quantum statistical 
averaging with respect to $H$. 
Within the strong coupling approach, when $H_t$ is treated as perturbation, the Dyson-type equation can be written in the matrix form as follows: 
\begin{equation}
\mathbf{G}_{ij}(\omega) = \bm{\Xi}_{ij}(\omega) + \sum_{\langle i'j'\rangle} \bm{\Xi}_{ij'}(\omega)  \cdot \frac{t_{j'i'}^{*}}{\sqrt{Z}} \mathbf{t} \cdot \mathbf{G}_{i'j}(\omega)\,,
\end{equation}
where $\bm{\Xi}_{ij}(\omega)$ is the irreducible cumulant~\cite{metzner:8549}  
which cannot be split into two disconnected parts by removing a single hopping line.

In the $Z\to\infty$ limit, the irreducible cumulant is local~\cite{metzner:8549}
\begin{equation}
\bm{\Xi}_{ij}(\omega) = \delta_{ij} \bm{\Xi}(\omega)
\end{equation}
and can be computed using the dynamical mean field theory (DMFT)~\cite{georges:13,freericks:1333}. 

Introducing the unperturbed  DOS of the Bethe lattice, 
\begin{equation}\label{eq:dos}
\rho(\epsilon) = \frac{2}{\piup W^2}\sqrt{W^2-\epsilon^2} \,, 
\end{equation}
we write the DMFT equation in the matrix form as~\cite{shvaika:075101}
\begin{equation}   \label{eq:DMFT} 
\mathbf{G}_{\text{local}}(\omega)
\equiv\mathbf{G}_{ii}(\omega)=  \int\limits_{-\infty}^{+\infty}\rd\epsilon \rho(\epsilon)
\mathbf{G}_{\epsilon} (\omega) 
 = 
\left[\bm{\Xi}^{-1}(\omega) - \bm{\Lambda}(\omega)\right]^{-1} 
= \mathbf{G}_{\text{imp}}(\omega)\,,
\end{equation}
where $\bm{\Lambda}(\omega)=[\lambda^{\alpha\beta}(\omega)]$ is the dynamical mean field ($\lambda$-matrix),  
$\mathbf{G}_{\text{imp}}(\omega)$ is the Green's function for the auxiliary impurity problem, and 
\begin{align}
\mathbf{G}_{\epsilon} (\omega)&=\left[\bm{\Xi}^{-1}(\omega) - \mathbf{t}\epsilon\right]^{-1}	
\end{align}
is the lattice Green's function matrix with the components
\begin{align}
G_{\epsilon}^{\beta\alpha}(\omega) &= \frac{A_{\beta\alpha}(\omega) - B_{\beta\alpha}\epsilon}{C(\omega) - D(\omega)\epsilon + \epsilon^2 \det\mathbf{t}} .
\label{GkvsE}
\end{align}
Here, we have introduced quantities
\begin{equation}
\mathbf{A}(\omega) = \adj \bm{\Xi}^{-1}(\omega) = \bm{\Xi}(\omega)/\det\mathbf{\Xi}(\omega)
\end{equation}
and
\begin{equation}
\mathbf{B} = \adj \mathbf{t} = \mathbf{t}^{-1}\det\mathbf{t} \,.
\end{equation}
For the $2\times2$ matrices,  the scalars $C$ and $D$ are given by 
\begin{equation}
C(\omega) = \det \mathbf{A}(\omega) = \det \bm{\Xi}^{-1}(\omega) = 1/\det \bm{\Xi}(\omega)
\end{equation} 
and
\begin{align}
D(\omega) = \Tr \left[\mathbf{A}(\omega)\mathbf{t}\right] = \Tr \left[\bm{\Xi}^{-1}(\omega)\mathbf{B}\right].
\end{align}

For the Bethe lattice, we can write the DMFT equation \eqref{eq:DMFT} as~\cite{shvaika:075101}
\begin{equation}\label{eq:DMFT_Bethe}
\bm{\Lambda}(\omega)=\frac{W^2}{4}\mathbf{t}\mathbf{G}_{\text{imp}}(\omega)\mathbf{t}\,,
\end{equation}
where the Green's function of the impurity problem with correlated hopping for the Falicov-Kimball model
is given by an exact expression
\begin{equation}
G_{\text{imp}}^{++}(\omega)  = w_1 g_1(\omega)\,,
\qquad
G_{\text{imp}}^{--}(\omega)  = w_0 g_0(\omega)\,,
\qquad
G_{\text{imp}}^{+-}(\omega)  = G_{\text{imp}}^{-+}(\omega) = 0\,.
\label{eq:Gimp}
\end{equation}
Here $w_1=\langle P^+\rangle=\langle n_f\rangle$, $w_0=\langle P^-\rangle=\langle 1- n_f\rangle$, and
\begin{align}
g_0(\omega) & =	\frac{1}{\omega+\mu_d - \lambda^{--}(\omega)}\,,
\nonumber \\
g_1(\omega) & = \frac{1}{\omega+\mu_d - U - \lambda^{++}(\omega)}
\label{eq:gimp0}
\end{align}
are the impurity Green's functions of a conduction electron 
in the presence of a $f$-state which is either permanently empty or occupied 
(i.e., locators in the CPA theory~\cite{blackman:2412}).

Expressions~\eqref{eq:gimp0} enable us to write a system of equations
\begin{align}
&\omega+\mu_d-U-\frac{1}{g_1(\omega)}
=\frac{W^2}{4}\left[(\tpp)^2 w_1 g_1(\omega)+(\tpm)^2 w_0 g_0(\omega)\right] ,
\nonumber\\
&\omega+\mu_d-\frac{1}{g_0(\omega)}
=\frac{W^2}{4}\left[(\tpm)^2 w_1 g_1(\omega)+(\tmm)^2 w_0 g_0(\omega)\right] ,
\label{eq:g0system}
\end{align}
which, in general, provide the 4th order  polynomial equations for $g_0(\omega)$ or $g_1(\omega)$.
In general, the quartic polynomial equation with real coefficients has either four real roots, or two real and two mutually conjugated complex roots, or two pairs of mutually conjugated complex roots. The physical solution is the one with negative imaginary parts of the retarded Green's functions $g_0(\omega)$ and $g_1(\omega)$ and it can be shown that there is always just a single set of physical solutions. 
For the frequencies $\omega$ having only real roots, the solution is obtained by using the spectral relation
\begin{equation}
\Re g_{0,1}(\omega)=-\frac{1}{\piup}\int_{-\infty}^{+\infty} \rd\omega^{\prime} \frac{\Im g_{0,1}(\omega^{\prime})}{\omega-\omega^{\prime}}\,, 
\end{equation}
which yields the correct retarded Green's functions and renormalized single-particle density of states
\begin{equation}
A_d(\omega) = -\frac{1}{\piup} \sum_{\alpha,\beta=\pm} \Im G_{\text{imp}}^{\alpha\beta}(\omega)
= -\frac{1}{\piup} \left[w_0 \Im g_{0}(\omega) + w_1 \Im g_{1}(\omega)\right]  .
\end{equation}
In numerical calculations, we use $W=2$, which defines our energy scale. 

For a given value of the $d$-electron concentration, $n_d=\langle n_d\rangle$, the chemical potential $\mu_d$ is obtained by solving the equation
\begin{equation}
n_d = -\frac{1}{\piup} \int_{-\infty}^{+\infty} \rd\omega f(\omega) \Im G_{\text{imp}}(\omega)\,,
\end{equation}
where $f(\omega)=1/(\re^{\omega/T}+1)$ is the Fermi-Dirac distribution function.

\section{Optical conductivity in the presence of correlated hopping}\label{sec:optcond}

Now, we develop the formalism for optical conductivity for the systems with correlated hopping. The linear response optical conductivity is determined (via the Kubo-Greenwood formula~\cite{kubo:570,greenwood:585}) by the imaginary part of the analytic continuation of the current-current correlation function to the real axis
\begin{equation}\label{eq:sigma:def}
\sigma(\Omega) = \frac{1}{\Omega} \Im \chi(\Omega)\,,
\end{equation}
where the current-current correlation function is defined by
\begin{equation}\label{eq:chi:def}
\chi(\tau-\tau') = \langle\mathcal{T} j(\tau) j(\tau')\rangle.
\end{equation}
An expression for the current operator for the Falicov-Kimball model with correlated hopping was derived in~\cite{shvaika:43704} for the infinite dimensional hypercubic lattice.

In infinite dimensions, due to the odd parity of the current vertex, the vertex corrections to the  current-current correlation function vanish~\cite{khurana:1990,moeller:7427} and the corresponding DMFT expression, generalized to the case of correlated hopping, contains only a bare bubble term
\begin{align}
\Im \chi(\Omega)&= \frac{1}{\piup}  \int \rd\omega \left[f(\omega+\Omega)-f(\omega)\right]  \int \rd \epsilon \rho(\epsilon) \Phi_{xx}(\epsilon) \Tr \left[\mathbf{t}\, \Im \mathbf{G}_{\epsilon} (\omega)\,  \mathbf{t}\, \Im \mathbf{G}_{\epsilon}(\omega+\Omega)\right],
\nonumber \\
\Re \chi(\Omega)&= -\frac{1}{\piup}  \int \rd\omega \biggl\{ f(\omega) \int \rd \epsilon \rho(\epsilon) \Phi_{xx}(\epsilon) \Tr \left[\mathbf{t}\, \Im \mathbf{G}_{\epsilon} (\omega)\,  \mathbf{t}\, \Re \mathbf{G}_{\epsilon}(\omega+\Omega)\right]
\nonumber\\
&+f(\omega+\Omega) \int \rd \epsilon \rho(\epsilon) \Phi_{xx}(\epsilon) \Tr \left[\mathbf{t}\, \Re \mathbf{G}_{\epsilon}(\omega)\,  \mathbf{t}\, \Im \mathbf{G}_{\epsilon} (\omega+\Omega)\right]\biggr\},
\label{eq:chi:gen}
\end{align}
where $\Phi_{xx}(\epsilon)$  is the so-called lattice-specific transport DOS~\cite{arsenault:205109}.
For a $D=\infty$ hypercubic lattice with Gaussian DOS, we have 
$\Phi_{xx}(\epsilon)=W^2/2D$, whereas, for the $Z=\infty$ Bethe lattice with semielliptic DOS,  
the $f$-sum rule for optical conductivity, which in the case of the nearest neighbour hopping reads~\cite{millis:10807}
\begin{equation}\label{eq:fsumrule}
\int_{0}^{+\infty} \rd \Omega\; \sigma(\Omega) = -\frac{\piup}{2}\langle H_t\rangle
\end{equation}
with $H_t$ defined in the case of correlated hopping by equation~\eqref{eq:Ht_mtrx}, yields~\cite{chung:11955,chattopadhyay:10738}
\begin{equation}
\Phi_{xx}(\epsilon) = \frac{1}{3Z}\left(W^2-\epsilon^2\right).
\label{eq:Phi_Bethe}
\end{equation}
The integral over $\epsilon$ in equation~\eqref{eq:chi:gen} can be evaluated now and we find that 
the final result depends on the value of $\det \mathbf{t}$. 

For $\det\mathbf{t}=0$, we get
\begin{align} 
\Im \chi(\Omega)&=\frac{1}{2\piup} \Re \int \rd \omega 
\left[f(\omega)-f(\omega+\Omega)\right] 
\left\{\frac{\Psi[E(\omega+\Omega)]-\Psi[E(\omega)]}{E(\omega+\Omega)-E(\omega)}-\frac{\Psi^*[E(\omega+\Omega)]-\Psi[E(\omega)]}{E^*(\omega+\Omega)-E(\omega)}\right\}, 
\nonumber \\ 
\Re \chi(\Omega)&=-\frac{1}{2\piup} \Im \int \rd \omega \left\{ 
\left[f(\omega)+f(\omega+\Omega)\right] 
\frac{\Psi[E(\omega+\Omega)]-\Psi[E(\omega) ]}{E(\omega+\Omega)-E(\omega) }\right. 
\nonumber\\ 
&+\left.\left[f(\omega)-f(\omega+\Omega)\right] 
\frac{\Psi^*[E(\omega+\Omega)]-\Psi\left[E(\omega) 
	\right]}{E^*(\omega+\Omega)-E(\omega)}\right\}, 
\end{align} 
where $E(\omega)=\frac{C(\omega)}{D(\omega)}$ and
\begin{align}
\Psi(\zeta)&=\int \rd \epsilon \frac{\rho(\epsilon)}{\zeta - \epsilon} \Phi_{xx}(\epsilon) .
\end{align}
For the semielliptic DOS, we have
\begin{align}
\Psi(\zeta)&=\frac{1}{3}\left[(W^2 -\zeta^2) F(\zeta)+ \zeta \right] ,
\end{align}
where the Hilbert transform of unperturbed DOS reads
\begin{align}
F(\zeta) &= \int \rd \epsilon \frac{\rho(\epsilon)}{\zeta - \epsilon}= \frac{2}{W^2}\left(\zeta - \sqrt{\zeta^2 -W^2}\right) .
\end{align}

For $\det\mathbf{t}\neq0$, we get for the imaginary part of current-current correlation function
\begin{align}
\Im \chi(\Omega)&=\frac{1}{2\piup (\det\mathbf{t})^2} \int \rd \omega \left[f(\omega+\Omega)-f(\omega)\right] \nonumber\\
&\times\Re  \left\{\frac{\Delta[E_1(\omega)]\Psi[E_1(\omega)]}{[E_2(\omega)-E_1(\omega)][E_1(\omega+\Omega)-E_1(\omega)][E_2(\omega+\Omega)-E_1(\omega)]}\right.
\nonumber\\
&+\frac{\Delta[E_2(\omega)]\Psi[E_2(\omega)]}{[E_1(\omega)-E_2(\omega)][E_1(\omega+\Omega)-E_2(\omega)][E_2(\omega+\Omega)-E_2(\omega)]}
\nonumber\\ &+\frac{\Delta[E_1(\omega+\Omega)]\Psi[E_1(\omega+\Omega)]}{[E_1(\omega)-E_1(\omega+\Omega)][E_2(\omega)-E_1(\omega+\Omega)][E_2(\omega+\Omega)-E_1(\omega+\Omega)]}
\nonumber\\
&+\frac{\Delta[E_2(\omega+\Omega)]\Psi[E_2(\omega+\Omega)]}{[E_1(\omega)-E_2(\omega+\Omega)][E_2(\omega)-E_2(\omega+\Omega)][E_1(\omega+\Omega)-E_2(\omega+\Omega)]}
\nonumber\\
&-\frac{\Delta^{\prime}[E_1(\omega)]\Psi[E_1(\omega)]}{[E_2(\omega)-E_1(\omega)][E^*_1(\omega+\Omega)-E_1(\omega)][E^*_2(\omega+\Omega)-E_1(\omega)]}
\nonumber\\
&-\frac{\Delta^{\prime}[E_2(\omega)]\Psi[E_2(\omega)]}{[E_1(\omega)-E_2(\omega)][E^*_1(\omega+\Omega)-E_2(\omega)][E^*_2(\omega+\Omega)-E_2(\omega)]}
\nonumber\\
&-\frac{\Delta^{\prime}[E^*_1(\omega+\Omega)]\Psi[E^*_1(\omega+\Omega)]}{[E_1(\omega)-E^*_1(\omega+\Omega)][E_2(\omega)-E^*_1(\omega+\Omega)][E^*_2(\omega+\Omega)-E^*_1(\omega+\Omega)]}
\nonumber\\ &\left.-\frac{\Delta^{\prime}[E^*_2(\omega+\Omega)]\Psi[E^*_2(\omega+\Omega)]}{[E_1(\omega)-E^*_2(\omega+\Omega)][E_2(\omega)-E^*_2(\omega+\Omega)][E^*_1(\omega+\Omega)-E^*_2(\omega+\Omega)]}
\right\},
\end{align}	
whereas for the real part we have
\begin{align}
\Re \chi(\Omega)&=\frac{1}{2\piup (\det\mathbf{t})^2} \int \rd \omega \biggl( \left[f(\omega)+f(\omega+\Omega)\right] 
\nonumber\\
&\times\Im  \left\{\frac{\Delta[E_1(\omega)]\Psi[E_1(\omega)]}{[E_2(\omega)-E_1(\omega)][E_1(\omega+\Omega)-E_1(\omega)][E_2(\omega+\Omega)-E_1(\omega)]}\right.
\nonumber\\ 
&+\frac{\Delta[E_2(\omega)]\Psi[E_2(\omega)]}{[E_1(\omega)-E_2(\omega)][E_1(\omega+\Omega)-E_2(\omega)][E_2(\omega+\Omega)-E_2(\omega)]}
\nonumber\\ 
&+\frac{\Delta[E_1(\omega+\Omega)]\Psi[E_1(\omega+\Omega)]}{[E_1(\omega)-E_1(\omega+\Omega)][E_2(\omega)-E_1(\omega+\Omega)][E_2(\omega+\Omega)-E_1(\omega+\Omega)]}
\nonumber\\
&\left.+\frac{\Delta[E_2(\omega+\Omega)]\Psi[E_2(\omega+\Omega)]}{[E_1(\omega)-E_2(\omega+\Omega)][E_2(\omega)-E_2(\omega+\Omega)][E_1(\omega+\Omega)-E_2(\omega+\Omega)]}\right\}
\nonumber\\
&-\left[f(\omega)-f(\omega+\Omega)\right] 
\Im \left\{\frac{\Delta^{\prime}[E_1(\omega)]\Psi[E_1(\omega)]}{[E_2(\omega)-E_1(\omega)][E^*_1(\omega+\Omega)-E_1(\omega)][E^*_2(\omega+\Omega)-E_1(\omega)]}\right. \nonumber\\
&+\frac{\Delta^{\prime}[E_2(\omega)]\Psi[E_2(\omega)]}{[E_1(\omega)-E_2(\omega)][E^*_1(\omega+\Omega)-E_2(\omega)][E^*_2(\omega+\Omega)-E_2(\omega)]} \nonumber\\
&+\frac{\Delta^{\prime}[E^*_1(\omega+\Omega)]\Psi[E^*_1(\omega+\Omega)]}{[E_1(\omega)-E^*_1(\omega+\Omega)][E_2(\omega)-E^*_1(\omega+\Omega)][E^*_2(\omega+\Omega)-E^*_1(\omega+\Omega)]} \nonumber\\ &\left.+\frac{\Delta^{\prime}[E^*_2(\omega+\Omega)]\Psi[E^*_2(\omega+\Omega)]}{[E_1(\omega)-E^*_2(\omega+\Omega)][E_2(\omega)-E^*_2(\omega+\Omega)][E^*_1(\omega+\Omega)-E^*_2(\omega+\Omega)]}\right\}\biggr)\,,
\end{align}	
where $E_{1}$ and $E_{2}$ are the roots of denominator in equation~\eqref{GkvsE}, 
$C(\omega) - D(\omega)\epsilon + \epsilon^2 \det\mathbf{t}=0$, given by  
\begin{align}
E_{1}(\omega) &= \frac{D(\omega)}{2\det \mathbf{t}} \left[ 1 + \sqrt{1 - \frac{4C(\omega)}{D^2(\omega)}\det \mathbf{t}}\, \right],
\label{eq:E1}
\\
E_{2}(\omega) &= \frac{2C(\omega)}{D(\omega)} \left[ 1 + \sqrt{1 - \frac{4C(\omega)}{D^2(\omega)}\det \mathbf{t}}\, \right]^{-1},
\label{eq:E2}
\end{align}
and $\Delta(\epsilon)$, $\Delta^{\prime}(\epsilon) $ reads 
\begin{align}
\Delta(\epsilon)&=[D(\omega)-\epsilon \det\mathbf{t}][D(\omega+\Omega)-\epsilon \det\mathbf{t}]+\epsilon^2 (\det\mathbf{t})^2-\det\mathbf{t} \Tr[\mathbf{A}(\omega)\bm{\Xi}^{-1}(\omega+\Omega)]\,,
\nonumber\\
\Delta^{\prime}(\epsilon)&=[D(\omega)-\epsilon \det\mathbf{t}][D^*(\omega+\Omega)-\epsilon \det\mathbf{t}]+\epsilon^2 (\det\mathbf{t})^2-\det\mathbf{t} \Tr[\mathbf{A}(\omega)\bm{\Xi}^{*-1}(\omega+\Omega)]\,.
\end{align}
The dc conductivity 
\begin{equation}\label{key}
\sigma_{\textrm{dc}} = \lim\limits_{\Omega\to0}\sigma(\Omega) = \sigma_0 \int_{-\infty}^{+\infty} \rd\omega \left[-\frac{d f(\omega)}{\rd\omega}\right] I(\omega)\,,
\end{equation}
on the one hand, can be derived from the above expressions and, on the other hand, it can be expressed in terms of the transport function $I(\omega)$. Expression for the transport function $I(\omega)$ and discussion of its anomalous properties are given in~\cite{dobushovskyi:125133}.

\section{Results and discussion}\label{sec:results}

\begin{figure}[!b]
	\centering
	\includegraphics[width=0.8\textwidth]{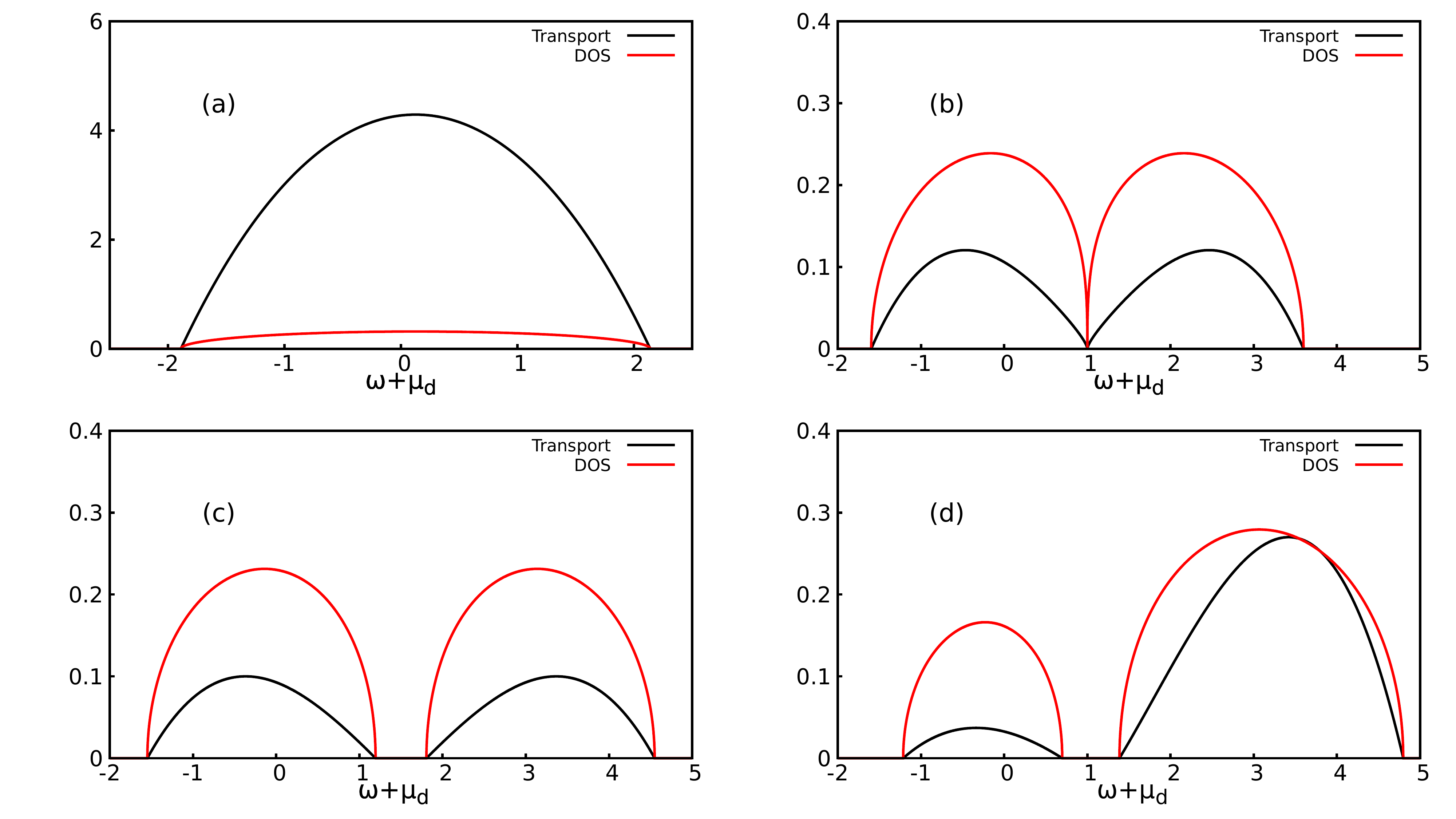}
	\caption{(Colour online) Density of states and transport function for (a) $U=0.25$ and (b) $U=2$ at $n_f=0.5$ and (c,d) $U=3$ at $n_f=0.5$ and $0.75$, respectively.} \label{fig:dos+tr_t2_0}
\end{figure}

First of all, we would like to recall the main features of the current-current correlation function $\chi(\Omega)$ and optical conductivity $\sigma(\Omega)$ for the systems without correlated hopping ($t_2=t_3=0$). In figure~\ref{fig:dos+tr_t2_0}, the density of states and transport function are presented for the metallic phase ($U=0.25$), critical Mott insulator ($U=2$), and strong Mott insulator ($U=3$) for different doping levels. 

\begin{figure}[!t]
	\centering
	\includegraphics[width=0.89\textwidth]{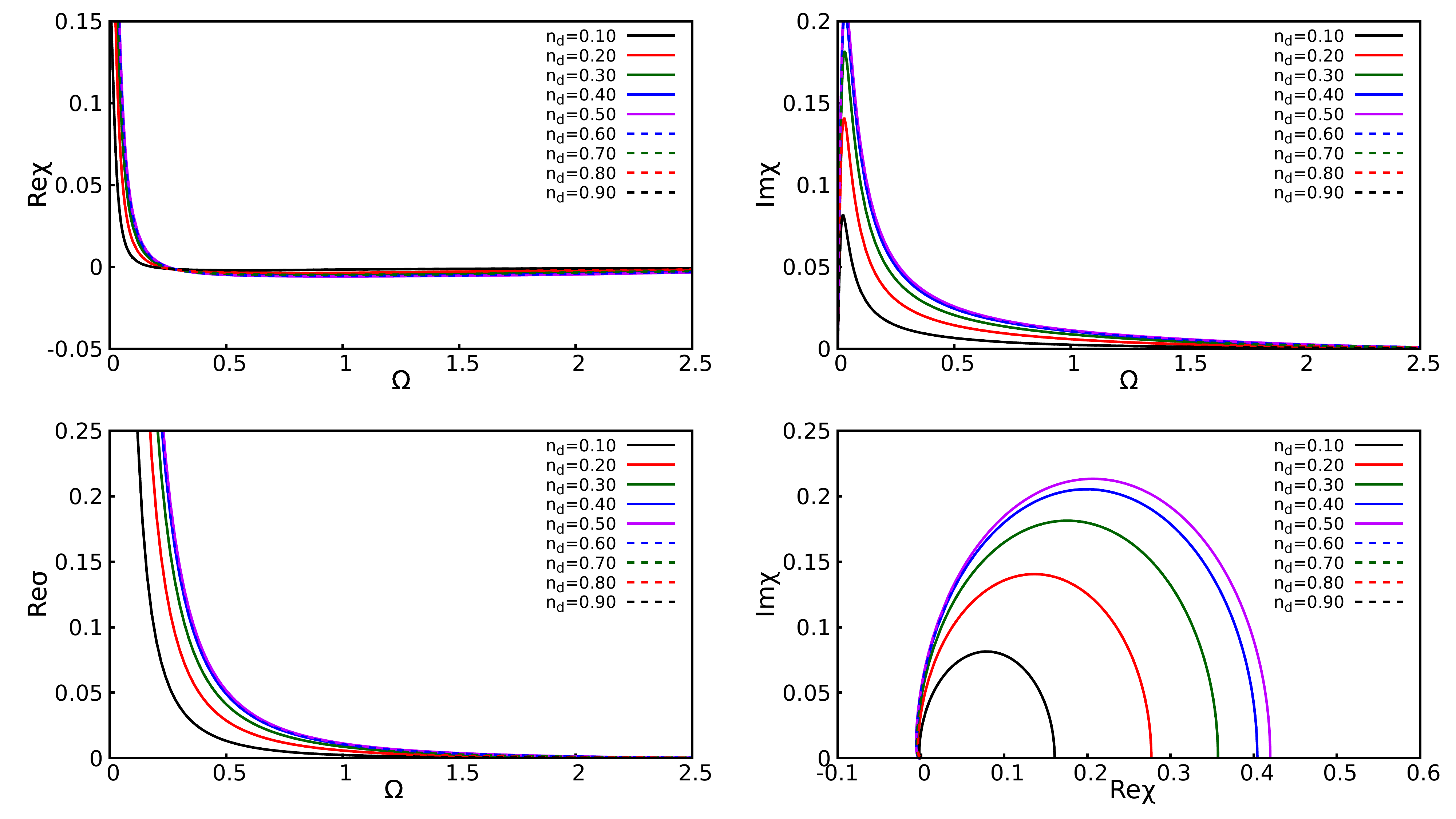}
	\caption{(Colour online) Current-current correlation function $\chi(\Omega)$, optical conductivity $\sigma(\Omega)$, and corresponding Nyquist plot for $U=0.25$, $n_f=0.5$, $T=0.05$, $t_2=t_3=0$.} \label{fig:U025}
	\bigskip
	\includegraphics[width=0.89\textwidth]{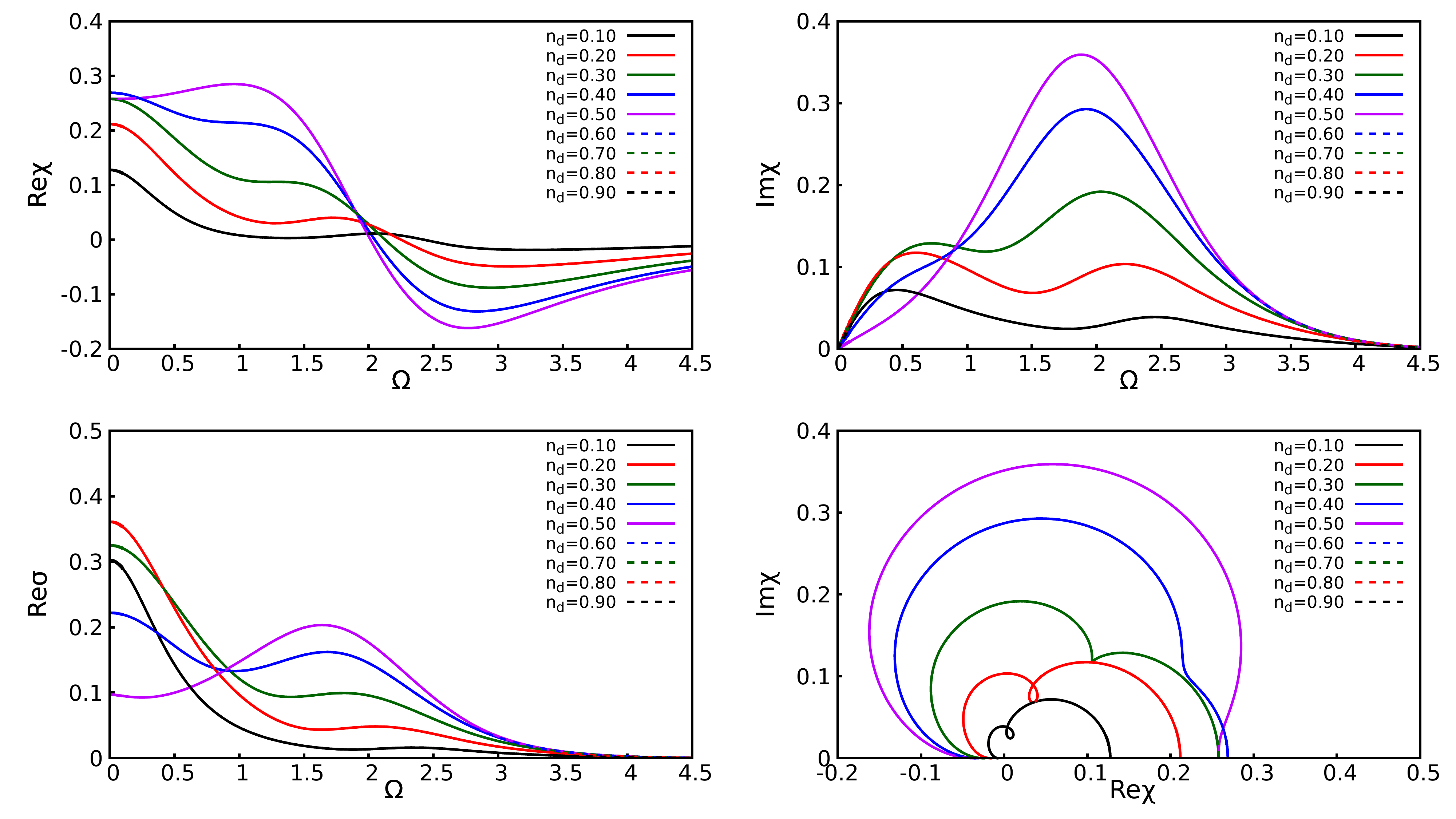}
	\caption{(Colour online) Current-current correlation function $\chi(\Omega)$, optical conductivity $\sigma(\Omega)$, and corresponding Nyquist plot for $U=2.0$, $n_f=0.5$, $T=0.15$, $t_2=t_3=0$.} \label{fig:U20}
\end{figure}

In the metallic phase, the DOS is smooth at the Fermi level [figure~\ref{fig:dos+tr_t2_0}(a)] and both the current-current correlation function $\chi(\Omega)$ and optical conductivity $\sigma(\Omega)$ display Drude peak at low frequencies, see figure~\ref{fig:U025}. The shape of Drude peak is described by the Debye relaxation equation
\begin{equation}\label{eq:Debye}
\chi_{\text{D}}(\Omega) = \chi_{\infty} + \frac{\chi_0- \chi_{\infty}}{1-\ri\Omega\tau_{\text{D}}},
\end{equation}
which is manifested by the semicircle on the Nyquist (Cole-Cole) plot for current-current correlation function. At large frequencies, there are deviations from the Debye relaxation and the semicircle in Nyquist plot is distorted in the vicinity of zero point.

For the critical Mott insulator ($U=2.0$) both DOS and transport functions possess a zero-width gap (pseudogap), see figure~\ref{fig:dos+tr_t2_0}(b). At half filling with $n_d=0.5$, the chemical potential is placed in the pseudogap, the dc conductivity is strongly reduced, and both the current-current correlation function $\chi(\Omega)$ and optical conductivity $\sigma(\Omega)$ do not display any Drude peak, whereas the charge-transfer peak at $\Omega\sim U$ is observed, see figure~\ref{fig:U20}. On doping $n_d\neq0.5$, the dc conductivity increases accompanied by the appearance of the Drude peak at low frequencies manifested as a semicircle segment on a Nyquist plot. Nevertheless, at high frequencies, the charge-transfer peak suppresses the Drude peak at a low doping level and gives a minor contribution at high doping levels. For the symmetric case $n_f=0.5$, the plots for the hole $n_d<1-n_f$ and electron $n_d>1-n_f$ doping are the same.
\vspace{-0.33mm}

\begin{figure}[!t]
	\centering
	\includegraphics[width=0.89\textwidth]{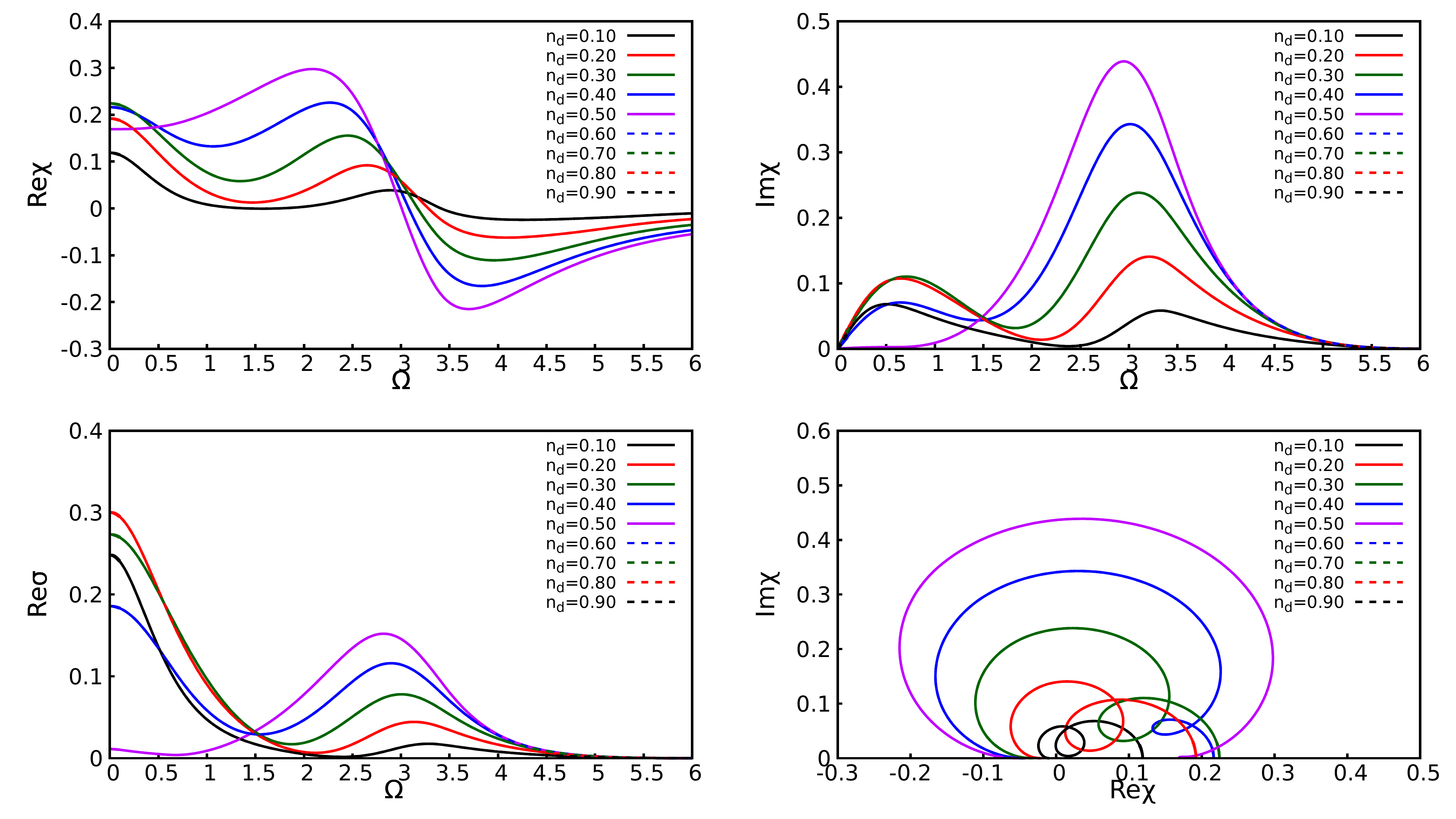}
	\caption{(Colour online) Current-current correlation function $\chi(\Omega)$, optical conductivity $\sigma(\Omega)$, and corresponding Nyquist plot for $U=3.0$, $n_f=0.5$, $T=0.15$, $t_2=t_3=0$.} \label{fig:U30}
	\bigskip
	\includegraphics[width=0.89\textwidth]{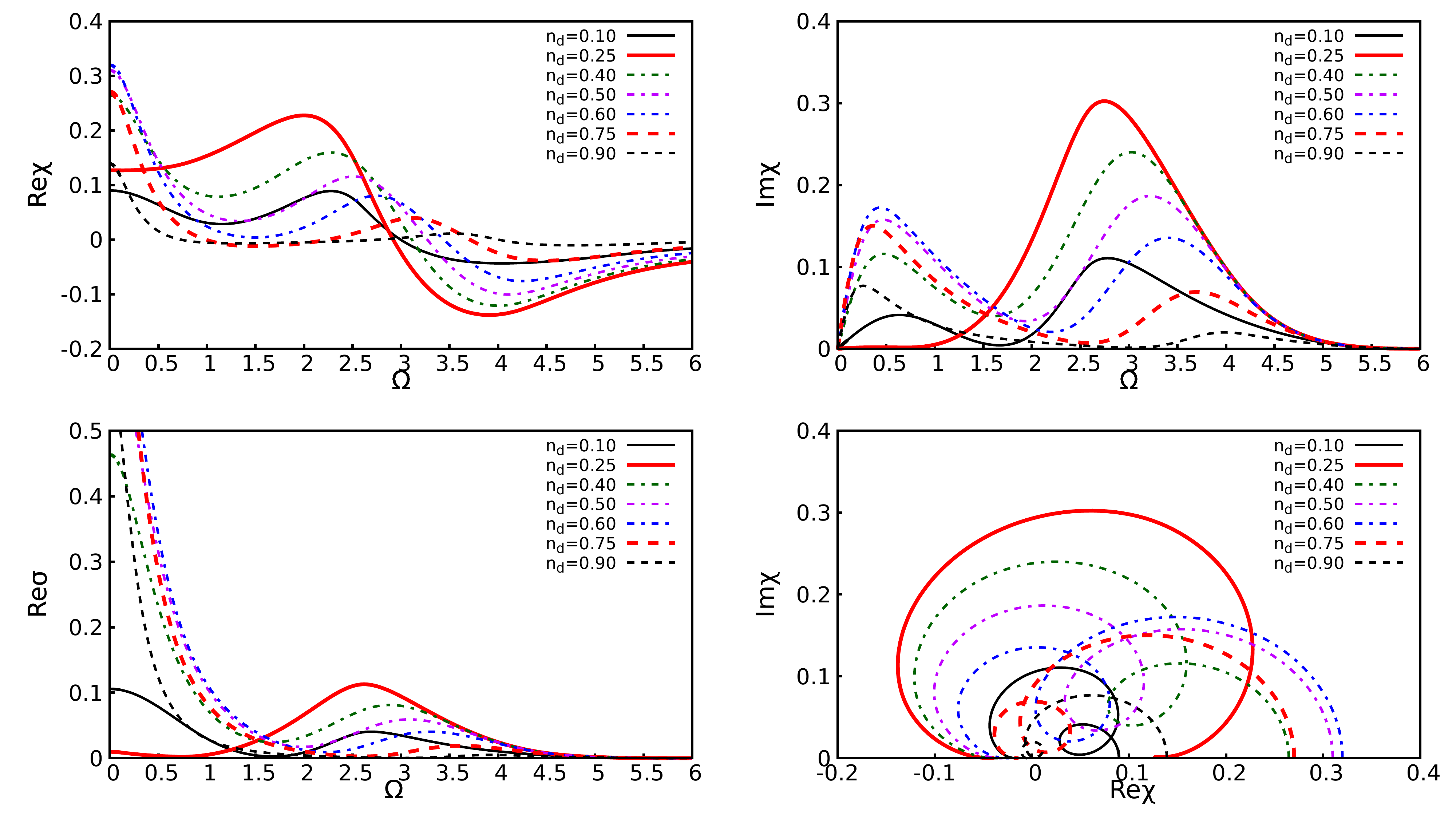}
	\caption{(Colour online) Current-current correlation function $\chi(\Omega)$, optical conductivity $\sigma(\Omega)$, and corresponding Nyquist plot for $U=3.0$, $n_f=0.75$, $T=0.15$, $t_2=t_3=0$.} \label{fig:U30nf075}
\end{figure}

For a larger interaction value $U=3$, deep in the Mott insulator phase, a large gap is observed on the DOS and transport function [figure~\ref{fig:dos+tr_t2_0}(c)] which governs the properties of the system. At half filling $n_d=n_f=0.5$ (figure~\ref{fig:U30}), both the current-current correlation function $\chi(\Omega)$ and the optical conductivity $\sigma(\Omega)$ contain only a charge-transfer peak which is represented by a circle on the Nyquist plot. The dc conductivity is completely suppressed and is of a thermal activation character. On doping, the dc conductivity is enhanced by the Drude peak which again manifests itself by a semicircle segment on the Nyquist plot. Nevertheless, the charge-transfer peak dominates the optical conductivity spectra.
	\vspace{-0.4mm}
	
Herein below, we shall consider the case of correlated hopping which is characterized by the asymmetric DOS, and it could be of interest to look at the asymmetric case for the ordinary Falicov-Kimball model without correlated hopping [figure~\ref{fig:dos+tr_t2_0}(d)]. In figure~\ref{fig:U30nf075}, we present the DOS and transport function for the case of $n_f=0.75$. Now, the Mott insulating case is observed at $n_d=1-n_f=0.25$ where the current-current correlation function $\chi(\Omega)$ as well as the optical conductivity $\sigma(\Omega)$ contain only a charge-transfer peak represented by a distorted circle on the Nyquist plot. Now, due to the asymmetry of DOS, the hole and electron doping displays different behaviour. For the hole doping ($n_d<1-n_f$), when the chemical potential is placed in the lower Hubbard band, the weak Drude peak  reappears but the spectrum is dominated by the charge-transfer peak. On the other hand, for the electron doping ($n_d>1-n_f$), the Drude peak arises very fast and becomes the major feature of the spectra.

\begin{figure}[!t]
	\centering
	\includegraphics[width=0.9\textwidth]{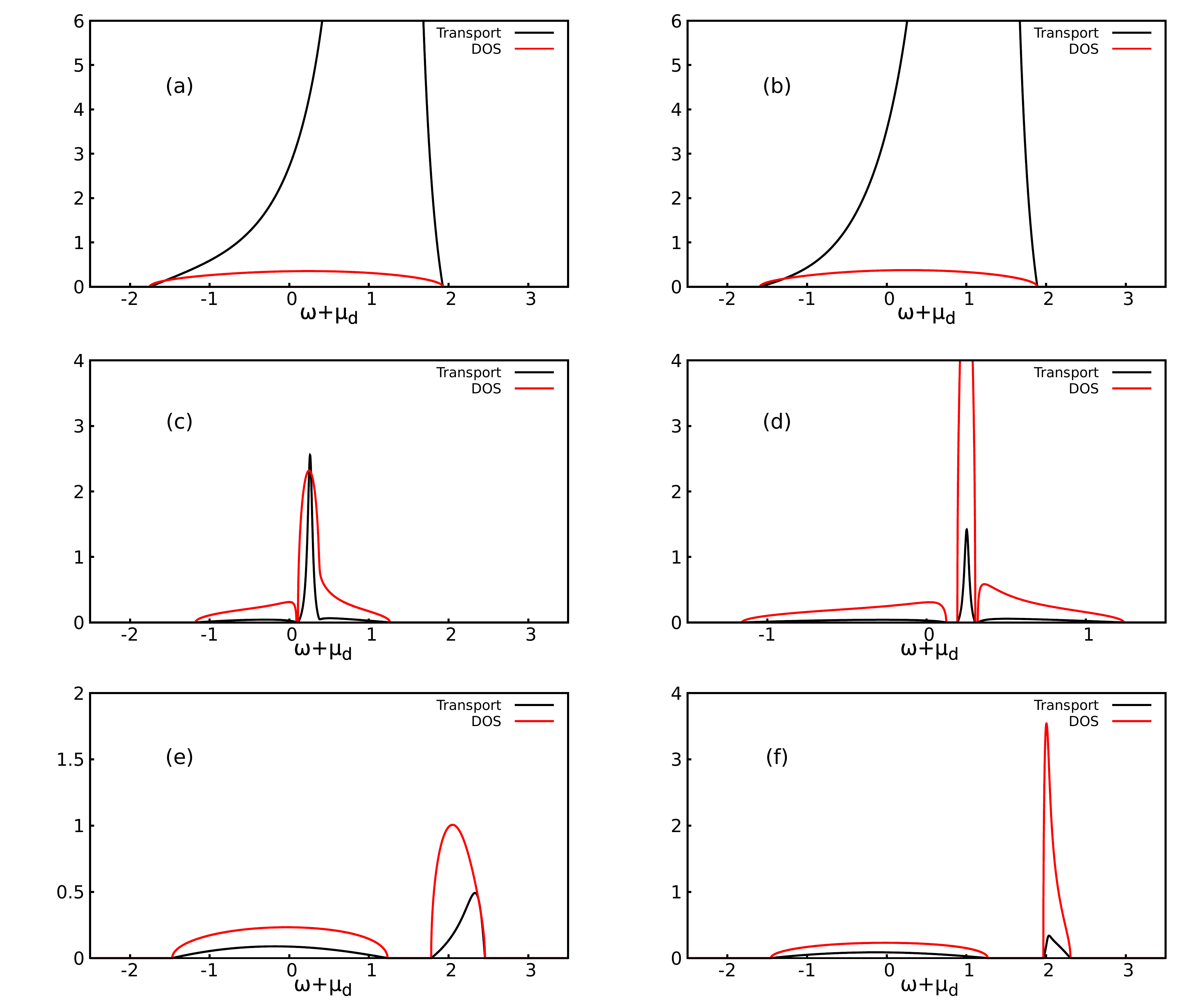} 
	\caption{(Colour online) Density of states and transport function for (a) $U=0.25$, $t_2=-0.1$, $n_f=0.5$, (b) $U=0.25$, $t_2=-0.1$, $n_f=0.75$, (c) $U=0.25$, $t_2=-0.45$, $n_f=0.75$, (d) $U=0.25$, $t_2=-0.48$, $n_f=0.75$, (e) $U=2$, $t_2=-0.4$, $n_f=0.5$, and (f) $U=2$, $t_2=-0.48$, $n_f=0.5$.} \label{fig:dos+tr}
\end{figure}

Now, let us consider the main features of the DOS and transport function in the presence of correlated hopping. In our previous article \cite{dobushovskyi:125133}, the following cases were distinguished. For small values of correlated hopping $t_2$ ($t_3=0$), the DOS becomes asymmetric and is slightly distorted in comparison with the one for the system without correlated hopping. On the other hand, the transport function $I(\omega)$, which determines the dc conductivity, is strongly enhanced by the resonant peak at $\omega=\omega_{\text{res}}$ [figure~\ref{fig:dos+tr}(a,b)], where
\begin{equation}
\omega_{\text{res}} + \mu_d = \frac{U}{1-\eta}
\label{eq:w_res}
\end{equation}
with
\begin{equation}\label{eq:x_res}
\eta = \frac{(t^{+-})^2}{(t^{--})^2} 
- \frac{(t^{+-})^2-\sqrt{(t^{+-})^4+4w_1w_0 \left[(t^{++}t^{--})^2-(t^{+-})^4\right] }}{2(t^{--})^2w_0}.
\end{equation}
A further increase of $t_2<0$ makes the bands narrow and lead to the opening of the gap in the spectrum for $t_2\approx -0.5$ ($t^{++}\approx0$) even for small values of $U$ [figure~\ref{fig:dos+tr}(c,d)]. For $n_f=0.5$, there are two bands with equal spectral weights of $0.5$, whereas for $n_f>0.5$, there are either two bands with spectral weights $1-n_f$ and $n_f$ for the lower and upper Hubbard bands, respectively, or three bands, two of which have equal spectral weights $1-n_f$ for the lower and upper Hubbard bands, while the third one with a spectral weight $2n_f-1$ originates from the clusters of the sites occupied by $f$-particles with a reduced hopping amplitude $t^{++}\approx0$. The resonant peak in now placed in the band of localized states. For a large interaction constant $U=2$, the band of localized states is observed only in a very narrow vicinity of the $t_2=-0.5$ value, and the DOS and transport functions contain two bands separated by a large gap.
\vspace{-0.5mm}
\begin{figure}[!t]
	\centering
	\includegraphics[width=0.9\textwidth]{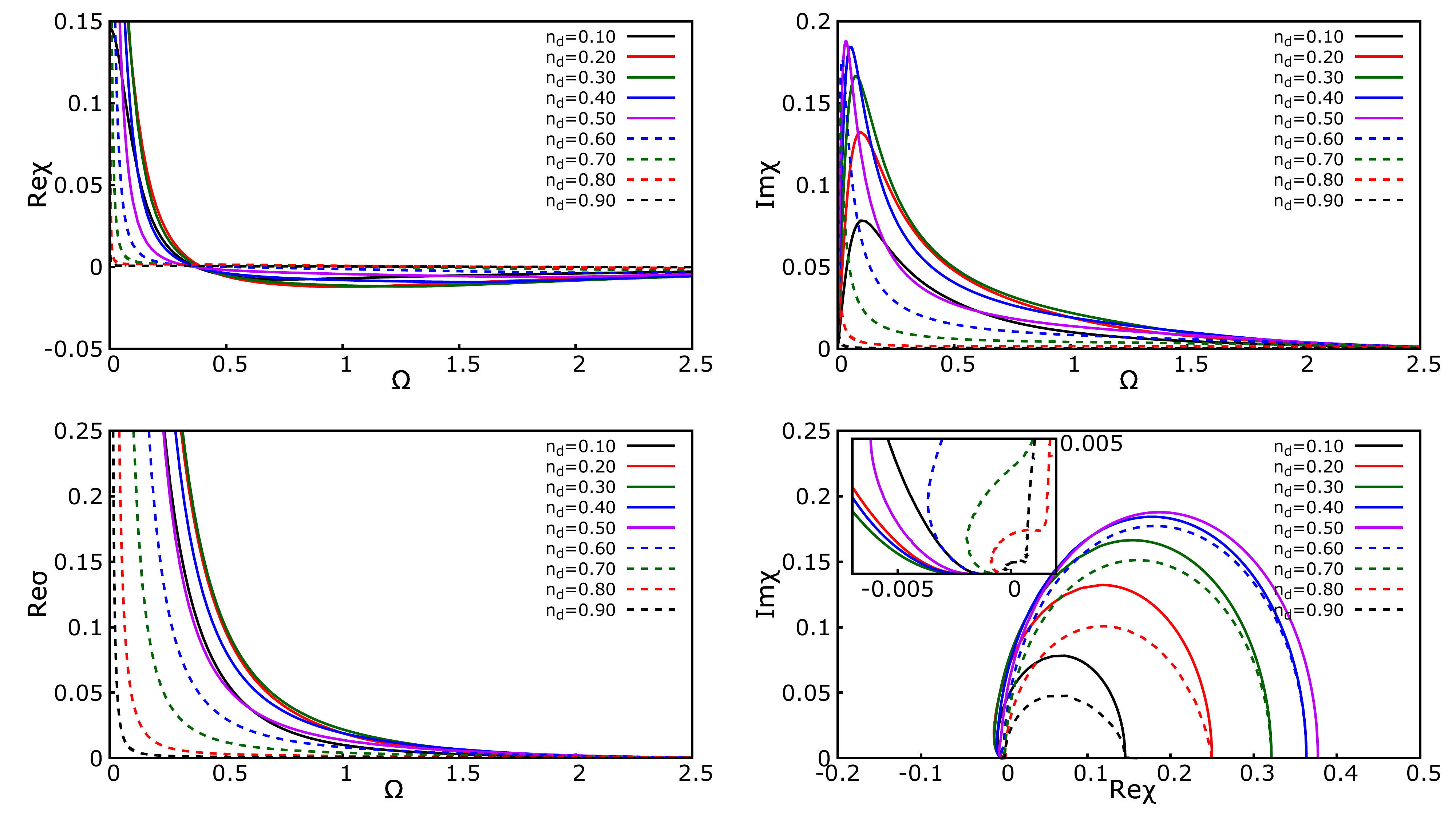}
	\caption{(Colour online) Current-current correlation function $\chi(\Omega)$, optical conductivity $\sigma(\Omega)$, and the corresponding Nyquist plot for $U=0.25$, $n_f=0.5$, $T=0.05$, $t_2=-0.1$, $t_3=0$.} \label{fig:U025t2_01}
	\bigskip
	\includegraphics[width=0.9\textwidth]{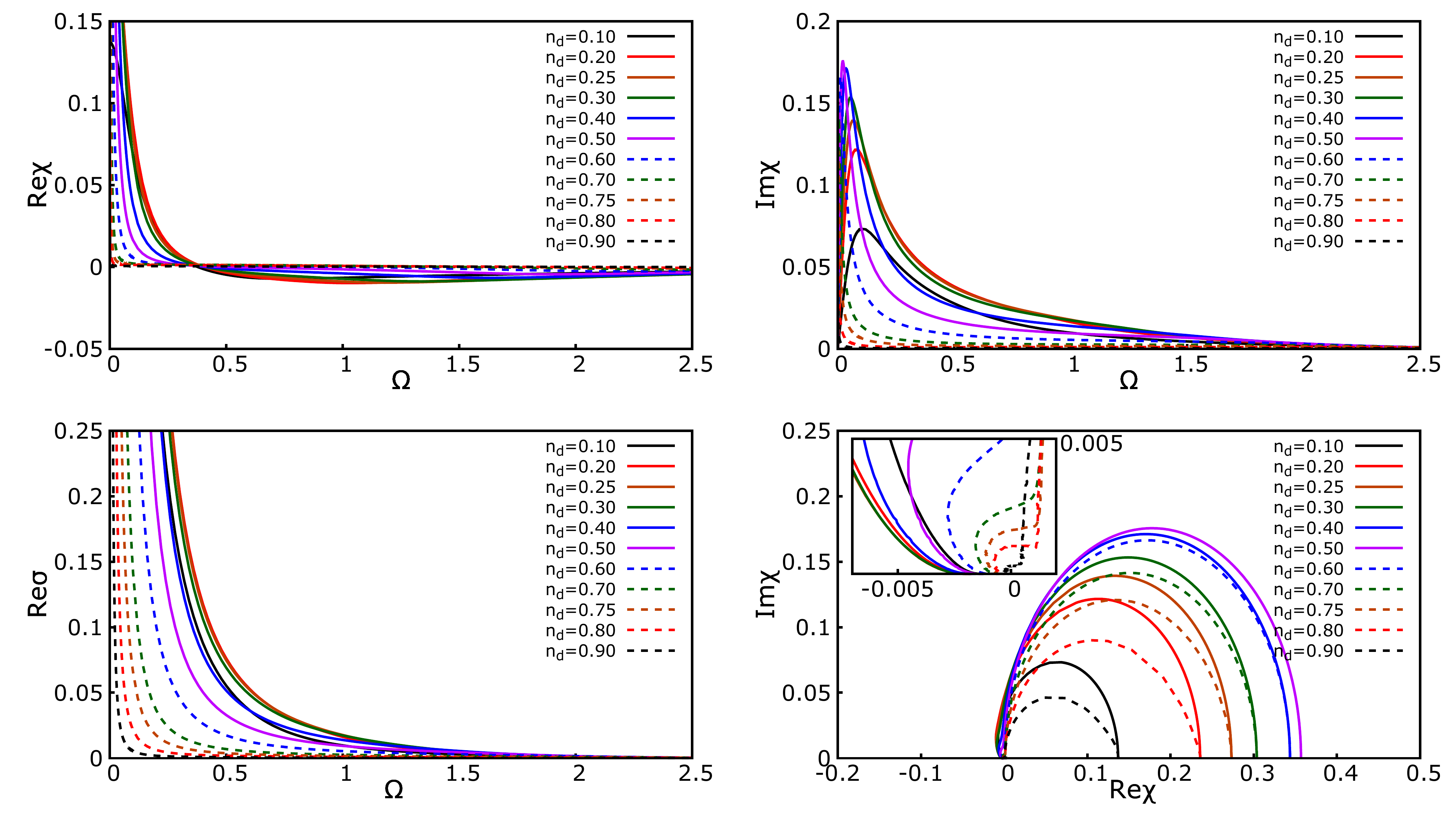}
	\caption{(Colour online) Current-current correlation function $\chi(\Omega)$, optical conductivity $\sigma(\Omega)$, and the corresponding Nyquist plot for $U=0.25$, $n_f=0.75$, $T=0.05$, $t_2=-0.1$, $t_3=0$.} \label{fig:U025t2_01nf075}
\end{figure}

For small values of correlated hopping, the shape of the current-current correlation function $\chi(\Omega)$ and optical conductivity $\sigma(\Omega)$ (figure~\ref{fig:U025t2_01}), at first glance, look the same as for the case without correlated hopping (figure~\ref{fig:U025}). For small doping levels, when chemical potential is far from the resonant peak, the Nyquist plots are almost semicircular and the most notable effect is a strong increase of the optical conductivity for low frequencies due to the resonant enhancement of a transport function. On the other hand, for a large doping level, when chemical potential is placed within the resonant peak, we observe a fast deviation of the Nyquist plots from semicircular shape at low frequencies, and the Debye relaxation equation~\eqref{eq:Debye} appears to be correct only for very small frequencies. Less noticeable is a growing up of the current-current correlation function at frequencies around $\Omega\sim1$, where an extended structure is observed. It is more observable as a bulge on the Nyquist plot in its left-hand part (see inserts in figures~\ref{fig:U025t2_01} and~\ref{fig:U025t2_01nf075}).
\vspace{-0.5mm}
\begin{figure}[!b]
	\centering
	\includegraphics[width=0.9\textwidth]{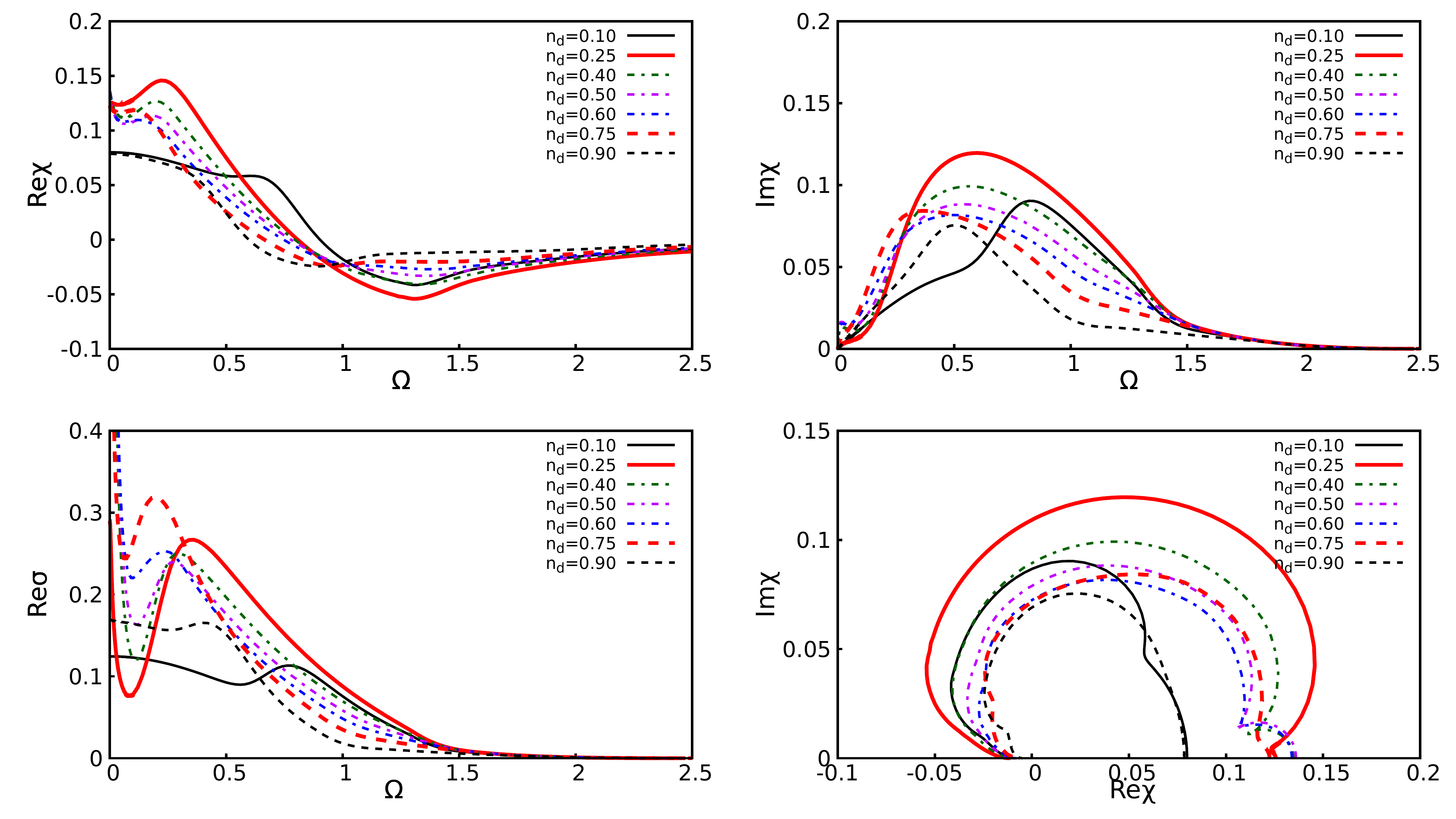}
	\caption{(Colour online) Current-current correlation function $\chi(\Omega)$, optical conductivity $\sigma(\Omega)$, and the corresponding Nyquist plot for $U=0.25$, $n_f=0.75$, $T=0.05$, $t_2=-0.45$, $t_3=0$.} \label{fig:U025t2_045}
	\bigskip
	\includegraphics[width=0.9\textwidth]{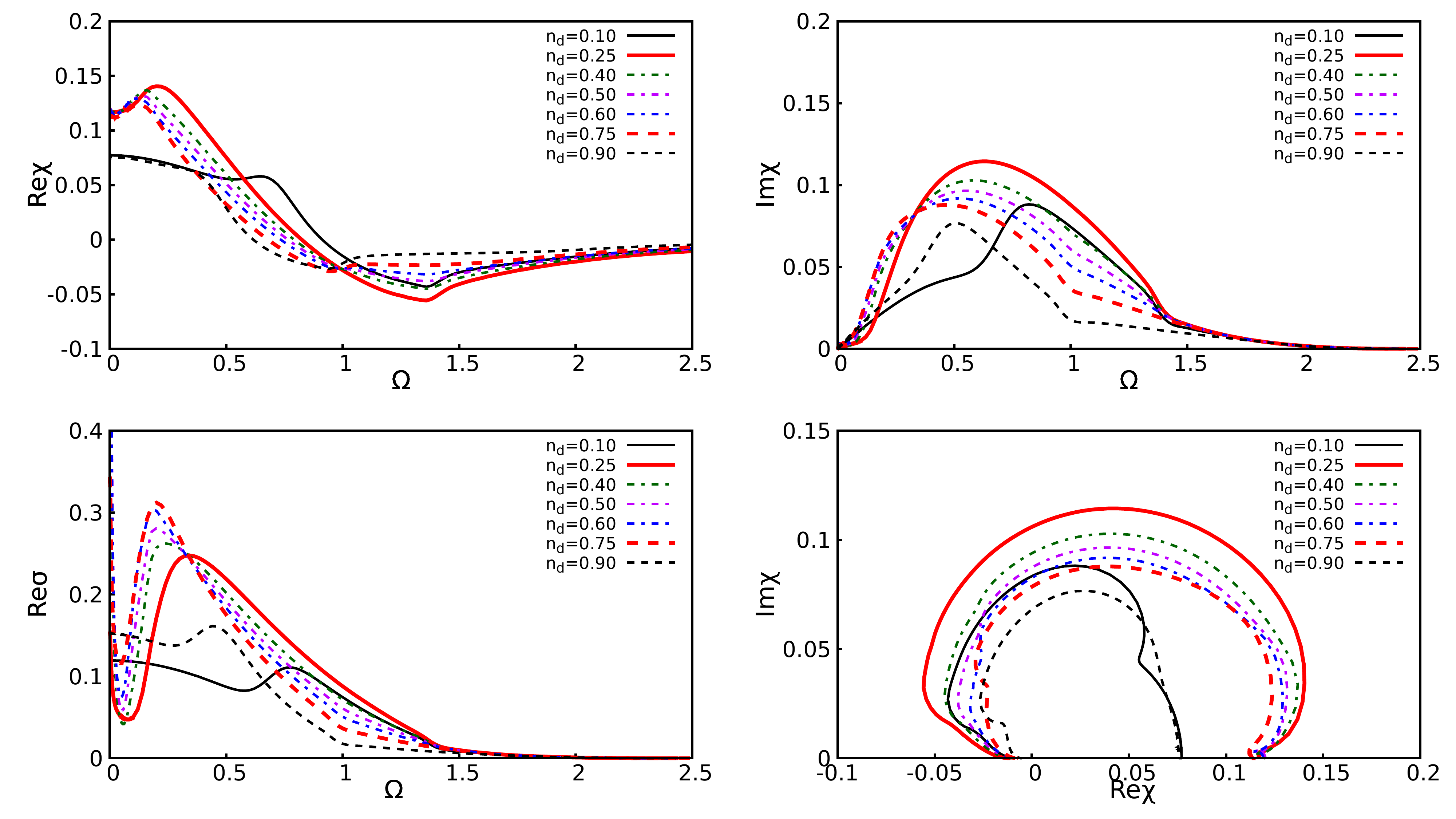}
	\caption{(Colour online) Current-current correlation function $\chi(\Omega)$, optical conductivity $\sigma(\Omega)$, and the corresponding Nyquist plot for $U=0.25$, $n_f=0.75$, $T=0.05$, $t_2=-0.48$, $t_3=0$.} \label{fig:U025t2_048}
\end{figure}

For the larger values of correlated hopping $t_2=-0.45$ and $-0.48$, when the gaps on DOS start to develop [see figure~\ref{fig:dos+tr}(c,d)], the shape of the current-current correlation function $\chi(\Omega)$ and optical conductivity $\sigma(\Omega)$ becomes very complicated (figures~\ref{fig:U025t2_045} and \ref{fig:U025t2_048}). One can distinguish two, three or four peaks depending on the temperature and placement of chemical potential (the peaks are wide and overlap which makes it difficult to recognize them). For $n_d<1-n_f$, when chemical potential is placed in the lower Hubbard band, there are at least four peaks. For a small gap insulator $n_d=1-n_f$, when the chemical potential is placed in the gap between the lower Hubbard band and the band of localized states, there are three peaks. For $1-n_f<n_d<n_f$ and $n_d>n_f$, when the chemical potential is placed in the band of localized states and in the upper Hubbard band, respectively, there are four peaks. For $n_d=n_f$, when chemical potential is placed in the localization gap, again we have at least three peaks. The complexity of the spectra is most evident on the Nyquist plots, where, besides the Drude and charge-transfer peaks, other complicated features are noticeable.
\begin{figure}[!t]
	\centering
	\includegraphics[width=0.9\textwidth]{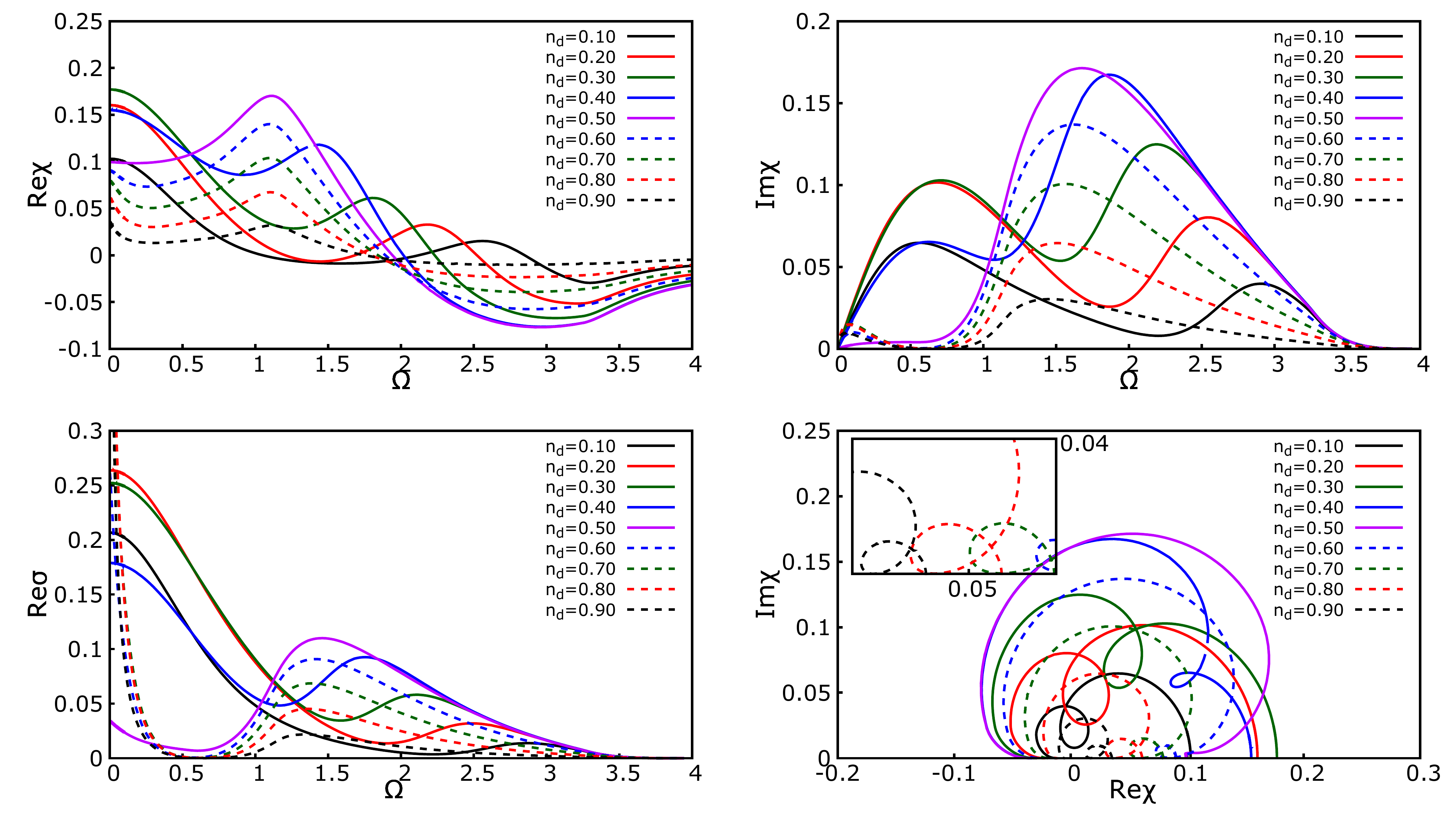}
	\vspace{-2mm}
	\caption{(Colour online) Current-current correlation function $\chi(\Omega)$, optical conductivity $\sigma(\Omega)$, and the corresponding Nyquist plot for $U=2.0$, $n_f=0.5$, $T=0.15$, $t_2=-0.4$, $t_3=0$.} \label{fig:U20t2_04}
\end{figure}
\begin{figure}[!b]
	\centering
	\includegraphics[width=0.9\textwidth]{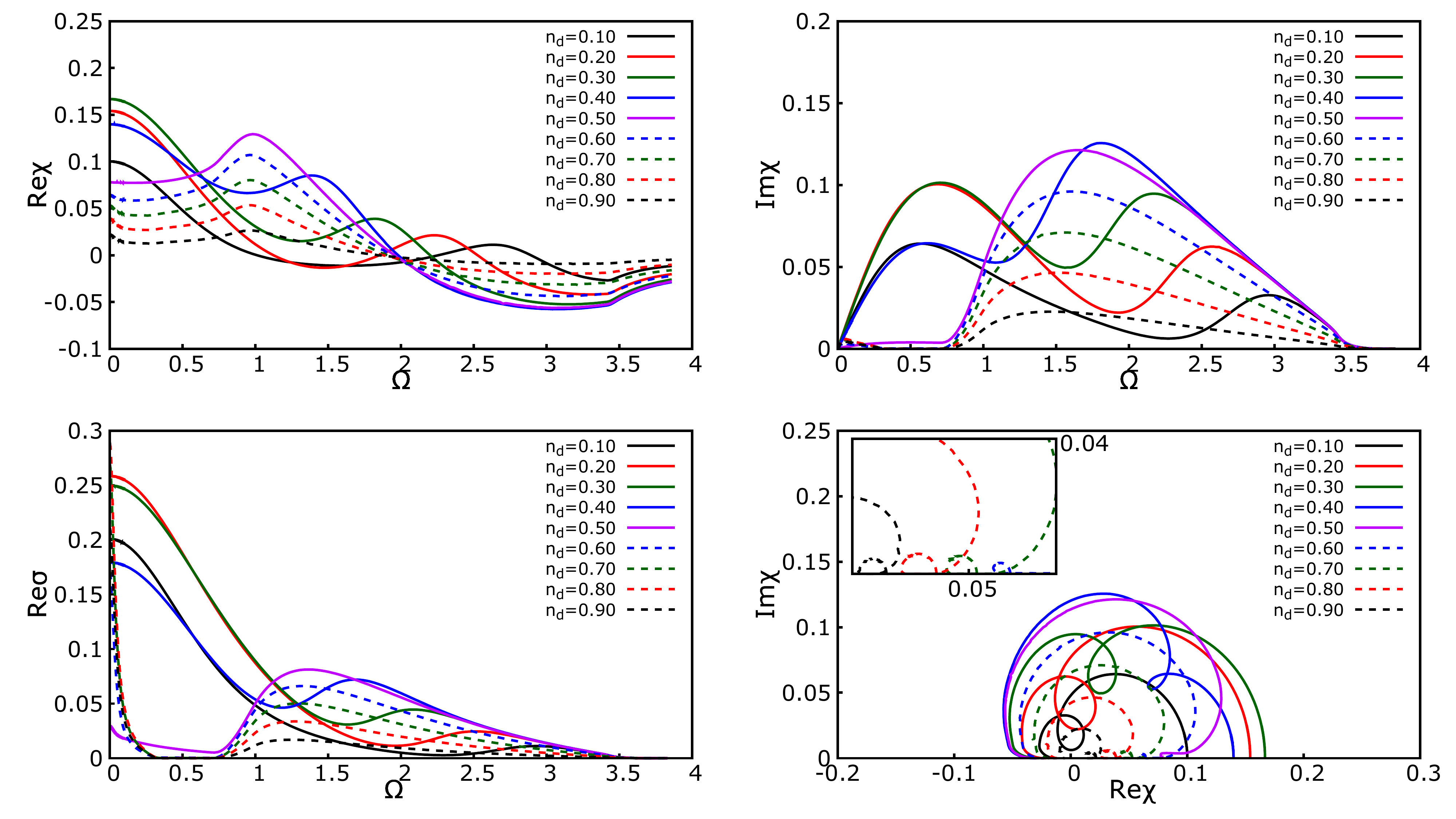}
	\vspace{-2mm}
	\caption{(Colour online) Current-current correlation function $\chi(\Omega)$, optical conductivity $\sigma(\Omega)$, and the corresponding Nyquist plot for $U=2.0$, $n_f=0.5$, $T=0.15$, $t_2=-0.48$, $t_3=0$.} \label{fig:U20t2_048}
\end{figure}

For the larger values of interaction constant $U=2$, there are two bands on DOS [figure~\ref{fig:dos+tr}(e,f)], the band of localization states is absorbed by the upper Hubbard band. Now, for a large doping level $n_d\geqslant1-n_f$, when the chemical potential is in the Mott gap or in the upper Hubbard band, the current-current correlation function $\chi(\Omega)$ and optical conductivity $\sigma(\Omega)$ (see figures~\ref{fig:U20t2_04} and \ref{fig:U20t2_048}) are dominated by the charge-transfer peak. On the other hand, in contrast to the asymmetric Mott insulator (figure~\ref{fig:U30nf075}), the Drude peak, which is now connected with an intraband transitions in a narrow upper Hubbard band, is also narrow and is separated from the charge-transfer peak by a gap. For a small doping level $n_d<1-n_f$, when the chemical potential is placed in a wide lower Hubbard band, the Drude peak is much wider and overlaps with the charge-transfer peak, and the total shape of the spectra is more similar to the case of a doped asymmetric Mott insulator. Nevertheless, in both cases, the low frequency behaviour of the current-current correlation function is well approximated by the Debye relaxation equation \eqref{eq:Debye}, which makes it possible to estimate the total spectral weight of the Drude peak on the optical conductivity
\begin{equation}
\int_{0}^{+\infty}\rd\Omega\, \frac{1}{\Omega} \Im\chi_{\text{D}}(\Omega) = \frac{\piup}{2} (\chi_0 - \chi_{\infty})
\end{equation}
as a diameter of semicircular segments on the Nyquist plots factor $\piup/2$. For low doping levels, the semicircles as well as the total spectral weight of the Drude peak are large (solid lines in figures~\ref{fig:U20t2_04} and~\ref{fig:U20t2_048}), whereas for large doping levels, the semicircles are small (dashed lines, see inserts) and the total spectral weight of the Drude peak is strongly reduced.

\section{Conclusions}\label{sec:conclusions}
\vspace{-1mm}

In this article we studied an influence of correlated hopping on the optical spectra for the Falicov-Kimball model on a Bethe lattice with a semielliptic DOS.

Using the dynamical mean field theory, we derived expressions for the current-current correlation function and computed an optical conductivity. Besides, the Nyquist plots were built and used to distinguish different contributions in the optical conductivity spectra.

For the metallic phase with a weak correlated hopping, when the transport function is dominated by the resonant peak, a strong enhancement of the Drude peak is developed. Besides, when chemical potential is placed in the vicinity of a resonant peak on transport function, a strong deviation from the Debye relaxation is observed at low frequencies, whereas at higher frequencies, a wide peak is observed on the optical conductivity, which is manifested by an additional circle on the Nyquist plots. 

For larger values of correlated hopping, an effective narrowing of the band-width with a simultaneous opening of a Mott-Hubbard gap and, at the finite doping, the emergence of a band of localized states is observed giving rise to a three-band structure of DOS. Now, the shape of optical conductivity spectra depends on the placement of the Fermi level. In the Mott insulator case, when the Fermi level is placed in a gap, which is small, the Drude peak is reduced, but not removed, and the main contribution is coming from the charge-transfer peak, which is strongly distorted due to correlated hopping. With doping, the Drude peak becomes more prominent, but its spectral weight is still much smaller than for the case without correlated hopping, and additional peaks on the optical conductivity arise due to the multi-band structure of DOS. For the case of strong local correlations, the overall picture depends on the doping level. For a small doping, when the chemical potential is placed in the wide lower Hubbard band, the spectral weight of the Drude peak is large, and the obtained results are much closer to the case of the doped Mott insulator without correlated hopping, whereas for a large doping, when the chemical potential is placed in the narrow upper Hubbard band, the spectral weight of the Drude peak is strongly reduced and it is separated by a gap from the charge-transfer peak.

\newpage

\ukrainianpart

\title{Нелокальні кореляції в спектрах оптичної провідності}
\author{Д.А. Добушовський, А.М. Швайка}
\address{
Інститут фізики конденсованих систем НАН України, вул. І. Свєнціцького, 1, 79011 Львів, Україна
}

\makeukrtitle

\begin{abstract}
\tolerance=3000%
Для моделі Фалікова-Кімбала з корельованим переносом на ґратці Бете досліджено спектри оптичної провідності. Використовуючи теорію динамічного середнього поля, отримано вирази для кореляційних функцій струм-струм. В металічній фазі і для малих значень корельованого переносу спостерігається відхилення форми піку Друде від дебаївської релаксації та виявлено появу додаткового широкого піку на спектрах оптичної провідності і діаграмах Найквіста, коли рівень Фермі знаходиться поблизу двочастинкового резонансу. Для більших значень величини корельованого переносу, густина станів містить три зони, а відповідні оптичні спектри та діаграми Найквіста набувають складної форми з додатковими піками. Для сильних локальних кореляцій врахування корельованого переносу приводить до звуження верхньої хаббардівської зони та зменшення спектральної ваги піку Друде для легованих моттівських діелектриків.
\keywords спектри оптичної провідності, модель Фалікова-Кімбала, корельований перенос, діаграма Найквіста

\end{abstract}

\end{document}